# Principles of the Electrical Resistance of Living Tissue

Maurice Philippson

*Abstract*—The impedance of one cubic centimeter of living tissues of potato and guinea pig were measured from 500 Hz to 3 MHz. In general, the magnitude of the impedance was found to monotonically decrease with increasing frequency. This implies that the membrane of each cell in the tissue acts like a capacitor, which is in parallel with a membrane resistance. The membrane resistance and capacitance together are in series with a protoplasm resistance. Experimentally, it was observed that after the guinea pig died, the membrane resistance of its muscle decreased from 1.49 kΩ to 0.79 kΩ while the protoplasm resistance remained around 0.11 kΩ. By contrast, when the potato started to germinate, the protoplasm resistance decreased from 0.25 kΩ to 0.10 kΩ, while its membrane resistance remained around 4 kΩ.

## I. Introduction

In a previous publication [1], we have outlined a method for producing alternating currents between DC and 3 MHz, and for measuring at these frequencies the electrical resistance of living tissues of plants or animals. These studies established the general principle of the electrical resistance of living tissues as a function of the frequency and certain tissue parameters.

The present experimental results (Figs. 1 to 3) show that, in general, the electrical resistance of tissues, measured in ohm/cm³, decreases monotonically from DC to infinite frequency. Therefore, the tissue resistance must be shunted by a capacitance $C$, so that its impedance decreases with the frequency $f$ as $1/2\pi fC$.

## II. Theory

We can experimentally demonstrate the existence of a capacitance in a living tissue. For example, let us take a cylinder of tissue from a potato with a measured resistance of 640 ohms at 100 kHz. Let us then take an inductive coil of about $10^5$ turns and 6 cm in diameter with a resistance of about 400 ohms at 100 kHz. Connecting the coil in series with the tissue cylinder, their total resistance becomes 380 ohms. This is because the magnitude of the impedance $Z$ of a resistance $R$ in series with an inductance $L$ and a capacitance $C$ is

$$|Z(\omega)| = \sqrt{R^2 + (\omega L - 1/\omega C)^2},$$

where $\omega = 2\pi f$ is the angular frequency. (The impedance of any network, however complicated, can be similarly simplified.) This shows that, when the addition of an inductance decreases the resistance of a network, the network must include a capacitance.

Based on the above observations, we hypothesize that the physiological membrane bounding a cell in the tissue can act as a capacitor. On the one hand, the membrane, essentially formed by lipids, resembles a perfect dielectric. On the other hand, the membrane is normally impermeable, not very permeable, or semi-permeable to electrolytes. (It is essential to distinguish the physiological membrane from the histological membrane. The former confines the cellular content and governs the physical process of absorption and the physiological process of permeability. The latter includes cellular secretion, which is part of the intercellular space. The thickness of the physiological membrane is on the order of a molecule.)

The cellular content and intercellular space on either side of the membrane conduct electric current and form the electrodes of the capacitor with the membrane as the dielectric. However, the electrical resistance is quite high for both the intracellular and intercellular media, since their main carrier of the electric current is a low-concentration solution of salt ions. This resistance is in series with the membrane capacitance.

In addition to the membrane capacitance, the membrane has a finite electrical resistance. If we consider a membrane across the entire tissue cylinder and perpendicular to the direction of the electric current, we find the membrane dielectric is interposed between the intercellular spaces. Therefore, based on what we just discussed, we must consider the membrane capacitance is shunted by a resistance.

It follows that the tissue can be represented electrically by a resistance $R$ in series with a capacitance $C$, with $C$ shunted by another resistance $r$. $R$ is the intrinsic electrolytic resistance of the cell protoplasm and the intercellular spaces of the tissue cylinder; $C$ and $r$ are the capacitance and resistance of the cell membranes in the tissue cylinder. The impedance $Z$ of such a network is given by

$$|Z(\omega)| = \sqrt{R^2 + \frac{2Rr + r^2}{1 + \omega^2 r^2 C^2}}. \quad (1)$$

At DC,

$$Z(0) = R + r, \quad (2)$$

Translated by James C. M. Hwang and Olivia Peytral-Rieu. J. C. M. Hwang is with the Department of Materials Science and Engineering, Cornell University, Ithaca, NY 148523, USA (e-mail: jch263@cornell.edu). O. P.-R. is with LAAS-CNRS, Toulouse University, 31400 Toulouse, France.

     

TABLE I
IMPEDANCE OF BLED GUINEA PIG LIVER
$R = 0.20$ k$\Omega$, $r = 1.78$ k$\Omega$, $C_1 = 0.59$ μF, $\alpha = 0.49$, $\beta = 0.2.7$ M$\Omega$

| Frequency (Hz) | Impedance Magnitude ($\Omega$) | | | | | Capacitance (F) | |
|---|---|---|---|---|---|---|---|
| | Measured | | | | Calculated | Calculated by (5) | Calculated by (7) |
| | 4 W | 12 W | 18 W | Average | | | |
| 500 | 1,950 | 1,700 | 2,200 | 1,950 | 1,957 | $3.2 \times 10^{-8}$ | $2.96 \times 10^{-8}$ |
| 1,000 | 1,900 | 1,715 | 2,250 | 1,948 | 1,934 | $1.63 \times 10^{-8}$ | $1.96 \times 10^{-8}$ |
| 2,000 | 1,900 | 1,695 | 2,180 | 1,925 | 1,913 | $1.1 \times 10^{-8}$ | $1.44 \times 10^{-8}$ |
| 5,000 | 1,700 | 1,625 | 1,965 | 1,763 | 1,775 | $9.25 \times 10^{-9}$ | $9.06 \times 10^{-9}$ |
| 10,000 | 1,625 | 1,510 | 1,800 | 1,645 | 1,610 | $5.75 \times 10^{-9}$ | $6.46 \times 10^{-9}$ |
| 20,000 | 1,490 | 1,345 | 1,585 | 1,475 | 1,390 | $3.8 \times 10^{-9}$ | $4.6 \times 10^{-9}$ |
| 50,000 | 1,070 | 875 | 1,025 | 990 | 1,036 | $3.15 \times 10^{-9}$ | $2.90 \times 10^{-9}$ |
| 100,000 | 800 | 680 | 855 | 780 | 800 | $2.17 \times 10^{-9}$ | $2.09 \times 10^{-9}$ |
| 200,000 | 615 | 525 | 639 | 595 | 598 | $1.5 \times 10^{-9}$ | $1.5 \times 10^{-9}$ |
| 500,000 | 440 | 382 | 447 | 425 | 417 | $9.25 \times 10^{-10}$ | $9.5 \times 10^{-10}$ |
| 1,000,000 | 327 | 296 | 351 | 325 | 326 | $6.9 \times 10^{-10}$ | $6.8 \times 10^{-10}$ |
| 1,250,000 | 300 | 285 | 325 | 303 | 300 | – | – |
| 1,500,000 | 280 | 274 | 300 | 285 | 290 | – | – |
| 2,000,000 | 270 | 260 | 294 | 275 | 270 | $4.4 \times 10^{-10}$ | $4.8 \times 10^{-10}$ |
| 2,500,000 | 248 | 249 | 272 | 255 | 255 | – | – |
| 3,000,000 | 235 | – | 255 | 245 | 242.5 | $4.16 \times 10^{-10}$ | $3.9 \times 10^{-10}$ |
| 10,000,000 | – | – | – | – | 215 | – | – |
| $\infty$ | – | – | – | – | 200 | – | – |

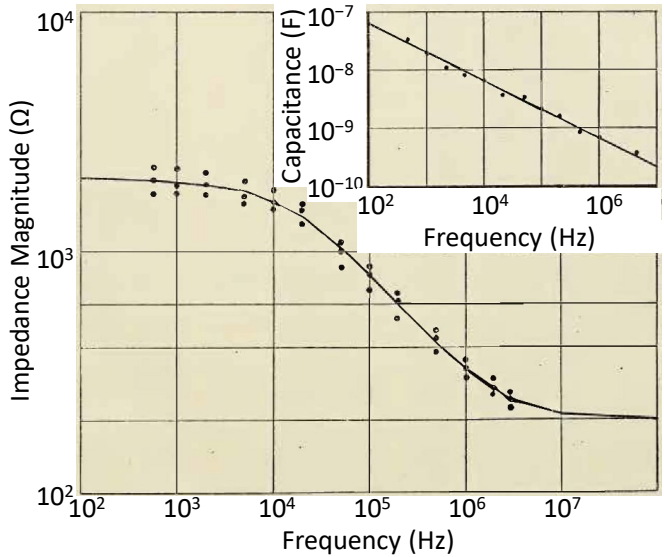

Fig. 1. Impedance of bled guinea pig liver.

which sets the upper limit of $Z$.

At the infinite frequency,

$$Z(\infty) = R, \quad (3)$$

which sets the lower limit of $Z$.

From (2) and (3),

$$r = Z(0) - Z(\infty). \quad (4)$$

Knowing the approximate values of $R$ and $r$, we can use (1) to calculate $C$ at each frequency,

$$C(\omega) = \frac{\sqrt{\dfrac{2Rr + r^2}{|Z|^2 - R^2} - 1}}{\omega r}. \quad (5)$$

Based on this calculation, we see that $C$ is not constant and its value decreases with frequency (Tables I to III). This should not surprise us. The capacitance of a capacitor is determined by its dimensions and the permittivity of the dielectric between the electrodes, and we know [2] that the permittivity of non-gaseous dielectrics decreases with frequency.

To study this phenomenon in the tissues, we plot in Figs. 1 to 3 the frequency dependence of the capacitance calculated by (5) from the measured impedance magnitude. With the capacitance and frequency both plotted on a logarithmic scale, it can be seen that the frequency dependence is indeed linear. Note that in these figures, circular symbols indicate measured impedance magnitudes and calculated capacitances by (5); solid curves indicate impedances calculated by (8) and capacitances calculated by (7). All measurements are performed at 25.5 °C on a cubic centimeter of tissue.

Apparently, the capacitance satisfies the following equation over a wide bandwidth:

$$\log C(\omega) = \log C(\omega') + \alpha \log \frac{\omega'}{\omega}. \quad (6)$$

The slope of the frequency dependence $\alpha$ can be determined graphically.

From (6), we have

$$C(\omega) \cdot \omega^\alpha = C(\omega') \cdot \omega'^\alpha = \text{constant}.$$

Let the constant be $C_1$, the capacitance at any frequency is given by

$$C(\omega) = C_1 \cdot \omega^{-\alpha}. \qquad (7)$$

TABLE II
IMPEDANCE OF BLED GUINEA PIG MUSCLE

| Frequency (Hz) | Immediately after Bleeding, $\alpha = 0.41$ | | | | 1 h after Bleeding, $\alpha = 0.41$ | | | |
|---|---|---|---|---|---|---|---|---|
| | $R = 0.11$ kΩ, $r = 1.49$ kΩ, $C_1 = 0.42$ μF, $\beta = 0.38$ MΩ | | | | $R = 0.11$ kΩ, $r = 0.79$ kΩ, $C_1 = 0.61$ μF, $\beta = 0.26$ MΩ | | | |
| | Impedance Magnitude (Ω) | | Capacitance (F) | | Impedance Magnitude (Ω) | | Capacitance (F) | |
| | Measured | Calculated | Calc. by (5) | Calc. by (7) | Measured | Calculated | Calc. by (5) | Calc. by (7) |
| 500 | 1,580 | 1,585 | $3.57 \times 10^{-8}$ | $3.3 \times 10^{-8}$ | 885 | 895 | $7 \times 10^{-8}$ | $4.8 \times 10^{-8}$ |
| 1,000 | 1,570 | 1,555 | $2.2 \times 10^{-8}$ | $2.56 \times 10^{-8}$ | 890 | 890 | $3 \times 10^{-8}$ | $3 \times 10^{-8}$ |
| 2,000 | 1,440 | 1,515 | $2.7 \times 10^{-8}$ | $1.88 \times 10^{-8}$ | 875 | 873 | $2.52 \times 10^{-8}$ | $2.68 \times 10^{-8}$ |
| 5,000 | 1,330 | 1,375 | $1.5 \times 10^{-8}$ | $1.28 \times 10^{-9}$ | 807 | 820 | $2 \times 10^{-8}$ | $1.84 \times 10^{-8}$ |
| 10,000 | 1,170 | 1,190 | $1.05 \times 10^{-8}$ | $9.55 \times 10^{-9}$ | 680 | 745 | $1.8 \times 10^{-8}$ | $1.38 \times 10^{-8}$ |
| 20,000 | 1,055 | 960 | $6.1 \times 10^{-9}$ | $7.2 \times 10^{-9}$ | 640 | 597 | $9.8 \times 10^{-9}$ | $1.04 \times 10^{-8}$ |
| 50,000 | 640 | 640 | $5.00 \times 10^{-9}$ | $5 \times 10^{-9}$ | 463 | 450 | $6.95 \times 10^{-9}$ | $7.16 \times 10^{-9}$ |
| 100,000 | 430 | 445 | $3.9 \times 10^{-9}$ | $3.74 \times 10^{-9}$ | 331 | 331 | $5.4 \times 10^{-9}$ | $5.4 \times 10^{-9}$ |
| 200,000 | 290 | 320 | $3.24 \times 10^{-9}$ | $2.8 \times 10^{-9}$ | 240 | 236 | $4.1 \times 10^{-9}$ | $4.06 \times 10^{-9}$ |
| 400,000 | 206 | 230 | $2.46 \times 10^{-9}$ | $2.97 \times 10^{-9}$ | 185 | 186 | $2.97 \times 10^{-9}$ | $3.05 \times 10^{-9}$ |
| 1,000,000 | 147 | 158 | $1.74 \times 10^{-9}$ | $1.45 \times 10^{-9}$ | 149 | 138 | $1.78 \times 10^{-9}$ | $2.09 \times 10^{-9}$ |
| 1,250,000 | 146 | 150 | $1.41 \times 10^{-9}$ | – | 138 | 133 | – | – |
| 1,500,000 | 143 | 142 | $1.23 \times 10^{-9}$ | $1.23 \times 10^{-9}$ | 132 | 128 | $1.64 \times 10^{-9}$ | $1.77 \times 10^{-9}$ |
| 2,000,000 | 139 | 134 | $1.05 \times 10^{-9}$ | $1.08 \times 10^{-9}$ | 130 | 123 | $1.3 \times 10^{-9}$ | $1.58 \times 10^{-9}$ |
| 2,500,000 | 131 | 129 | $1.24 \times 10^{-9}$ | $9.9 \times 10^{-10}$ | 124 | 121 | $1.2 \times 10^{-9}$ | $1.44 \times 10^{-9}$ |
| 3,000,000 | 129 | 125 | $9.05 \times 10^{-10}$ | $9.2 \times 10^{-10}$ | – | 118 | – | – |
| 10,000,000 | – | 113 | – | – | – | 112 | – | – |
| ∞ | – | 110 | – | – | – | 110 | – | – |

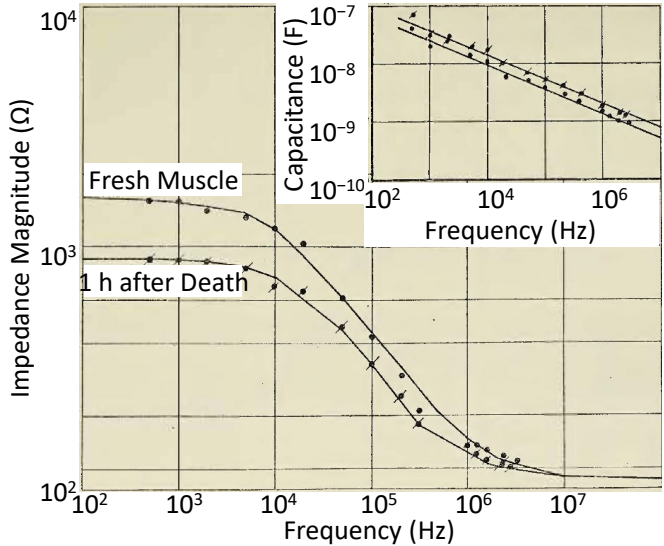

Fig. 2. Impedance of bled guinea pig muscle.

Substituting (7) in (1), we have

$$|Z(\omega)| = \sqrt{R^2 + \frac{2Rr + r^2}{1 + \omega^{2(1-\alpha)} r^2 C_1^2}}. \qquad (8)$$

Using (8), calculated values of $|Z|$ are listed in Tables I to III and plotted in Figs. 1 to 3. It can be seen that the calculated values are very close to the measured data.

Having experimentally validated our hypotheses, we can start to analyze the phenomenon we are studying. Indeed, in (7) if $\omega$ approaches zero (close to DC), $C$ approaches infinity. This means that the permittivity of the dielectric membrane approaches infinity. This implies the impedance of the membrane capacitance is a polarization reactance, which seems consistent with ongoing experiments on the impedance of copper-ferrocyanide membranes. In this case, the polarization reactance of a semipermeable membrane is

$$X_P = \frac{1}{j\omega^{(1-\alpha)} C_1}. \qquad (9)$$

The symbol $j$ denotes the square root of −1 and indicates $X_P$ is in phase with the current flowing through the capacitor.

Let

$$\frac{1}{C_1} = \beta,$$

then

$$X_P = \frac{\beta}{j\omega^{(1-\alpha)}}. \qquad (10)$$

The parameter $\beta$ in $\Omega/\text{cm}^3$ is the polarization reactance at the unity frequency. Substituting (10) into (8) gives

$$|Z(\omega)| = \sqrt{R^2 + \frac{2Rr + r^2}{1 + \omega^{2(1-\alpha)} \beta^2 r^2}}. \qquad (11)$$

In (11), $R$ in $\Omega/\text{cm}^3$ is the electrical resistance of the protoplasm and the intercellular spaces of the tissue under study. Expressed in resistivity, it is the inverse of conductivity. It indicates the state of electrolytes in the protoplasm. The parameter $r$ in $\Omega/\text{cm}^3$ is the ohmic resistance of the cell membranes. The parameter $\beta$ in $\Omega/\text{cm}^3$ is the polarization reactance of the membranes at the unity frequency. Higher $r$ and $\beta$ indicate a membrane less permeable to electrolytes. The numerical coefficient $\alpha$ indicates how fast the polarization reactance decreases with increasing frequency.

TABLE III
IMPEDANCE OF POTATO

| Frequency (Hz) | Dormant, 4 March 1921 | | | Germinated, mid-April, 1921 | | | | | |
|---|---|---|---|---|---|---|---|---|---|
| | $R = 0.25$ kΩ | $r = 4.33$ kΩ | $C_1 = 0.041$ μF | $R = 0.10$ kΩ | | $r = 4.66$ kΩ | | $C_1 = 0.041$ μF | |
| | $\alpha = 0.25$ | | $\beta = 3.55$ MΩ | $\alpha = 0.25$ | | | | $\beta = 3.55$ MΩ | |
| | Impedance Magnitude (Ω) | | Cap. (F) | Impedance Magnitude (Ω) | | | | Capacitance (F) | |
| | Measured | Calculated | Calc. by (5) | Measured | Measured | Average | Calculated | Calc. by (5) | Calc. by (7) |
| 500 | 4,535 | 4,540 | $1.44 \times 10^{-8}$ | 4,025 | 5,300 | 4,660 | 4,660 | $9 \times 10^{-9}$ | $9.5 \times 10^{-9}$ |
| 1,000 | 4,580 | 4,490 | $5.2 \times 10^{-9}$ | 4,100 | 5,250 | 4,675 | 4,550 | $3.6 \times 10^{-9}$ | $8 \times 10^{-9}$ |
| 2,000 | 4,400 | 4,290 | $4.76 \times 10^{-9}$ | 4,020 | 4,390 | 4,475 | 4,360 | $5.5 \times 10^{-9}$ | $6.7 \times 10^{-9}$ |
| 5,000 | 3,975 | 3,740 | $4.5 \times 10^{-9}$ | 3,260 | 3,740 | 3,500 | 3,720 | $6.5 \times 10^{-9}$ | $5.4 \times 10^{-9}$ |
| 10,000 | 3,420 | 2,920 | $3.27 \times 10^{-9}$ | 2,755 | 3,060 | 2,907 | 2,875 | $4.6 \times 10^{-9}$ | $4.7 \times 10^{-9}$ |
| 20,000 | 2,150 | 2,060 | $3.37 \times 10^{-9}$ | 2,275 | 2,490 | 2,382 | 1,965 | $3.4 \times 10^{-9}$ | $3.8 \times 10^{-9}$ |
| 50,000 | 1,130 | 1,150 | $2.94 \times 10^{-9}$ | 1,155 | 1,050 | 1,100 | 1,065 | $2.1 \times 10^{-9}$ | $3 \times 10^{-9}$ |
| 100,000 | 730 | 705 | $2.5 \times 10^{-9}$ | 664 | 604 | 634 | 664 | $2.7 \times 10^{-9}$ | $2.53 \times 10^{-9}$ |
| 200,000 | 500 | 448 | $1.95 \times 10^{-9}$ | 366 | 362 | 364 | 375 | $2.4 \times 10^{-9}$ | $2.15 \times 10^{-9}$ |
| 500,000 | 330 | 316 | $1.56 \times 10^{-10}$ | 202 | 206 | 204 | 213 | $1.85 \times 10^{-9}$ | $1.7 \times 10^{-9}$ |
| 1,000,000 | 265 | 275 | $1.91 \times 10^{-10}$ | 143 | 144 | 145 | 153 | $1.52 \times 10^{-9}$ | $1.42 \times 10^{-9}$ |
| 1,250,000 | 260 | 269 | $1.8 \times 10^{-9}$ | 129 | 131 | 130 | 138 | – | $1.34 \times 10^{-9}$ |
| 1,500,000 | 258 | 264 | $1.78 \times 10^{-9}$ | 118 | – | 118 | 132 | $1.73 \times 10^{-9}$ | $1.28 \times 10^{-9}$ |
| 2,000,000 | (250) | 260 | – | – | – | – | – | – | – |
| 2,500,000 | 257.5 | 259 | $1.08 \times 10^{-9}$ | – | – | – | – | – | $1.14 \times 10^{-9}$ |
| 10,000,000 | – | 251 | – | – | – | – | 101 | – | – |
| ∞ | – | 250 | – | – | – | – | 100 | – | – |

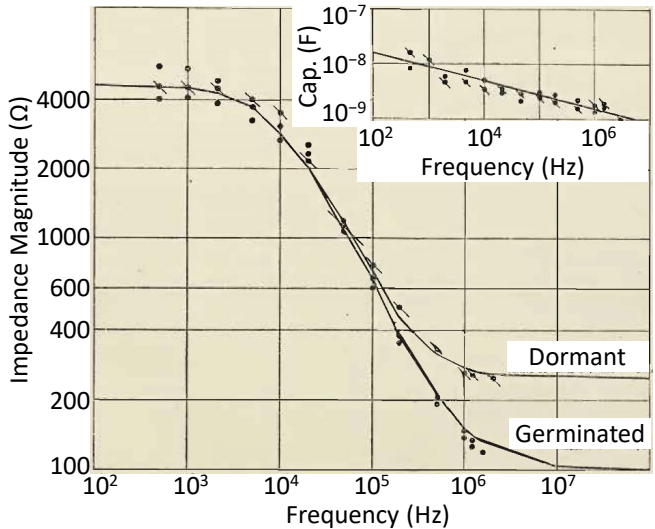

Fig. 3. Impedance of potato.

The question now is whether or not these parameters are fixed for a given tissue in a given physiological state, but variable with physiological changes.

### III. EXPERIMENT

The three series of measurements that we carried out on the liver of a guinea pig immediately after it is killed and bled agreed very well (Table I and Fig. 1). This demonstrates the stability of the tissue parameters and the precision of the measurement technique. By contrast, when we measured the muscle of the same animal, we found the tissue parameters changed rapidly resulting in mismatched data. To study these changes, we compare a series of measurements taken immediately after bleeding the animal (characteristic of the fresh tissue) with a second series of measurements made about an hour after the animal died (Table II and Fig. 2). The comparison shows that upon death the membrane impedance decreases noticeably, while the protoplasm resistance $R$ remains constant. Additionally, the frequency dependence of the tissue capacitance is the same for a cooler and denser tissue, indicating the parameter $\alpha$ remains constant while the membrane capacitance $C_1$ increases. This implies that the polarization reactance $\beta$ decreases along with the membrane resistance $r$.

Finally, we made two series of measurements on potatoes, in March during the winter dormancy and in April at the start of germination. This time, the results (Table III and Fig. 3) show a very large decrease in protoplasm resistance $R$, indicating the mobilization of the salt reserve, without changing membrane parameters such as $r$, $C_1$, $\alpha$, and $\beta$. In fact, the same straight line satisfies the calculated logarithmic capacitances for the two series of measurements. Therefore, we can use these parameters to analyze cell physiology, in particular, the link between the changes in these parameters and variations in the permeability of cell membranes.

### IV. CONCLUSION

The study of the electrical resistance of living tissues between 500 Hz and 3 MHz gave us the following:

1) In general, a cell membrane conducts alternating current like a capacitor.

2) The electrical conductivity of a living tissue can be expressed by the two following equations, which differ from each other only by the definition of the parameter $C_1$:

$$|Z(\omega)| = \sqrt{R^2 + \frac{2Rr + r^2}{1 + \omega^{2(1-\alpha)} r^2 C_1^2}} \ . \qquad (8)$$

$$|Z(\omega)| = \sqrt{R^2 + \frac{2Rr + r^2}{1 + \omega^{2(1-\alpha)} \beta^2 r^2}} \ . \qquad (11)$$

3) The polarization reactance of a cell membrane is analogous to that of a capacitor and is expressed by

$$X_P = \frac{\beta}{j\omega^{(1-\alpha)}} \ . \qquad (10)$$

In the above equations the symbols are defined as in the following:

$Z$ in $\Omega/\text{cm}^3$ is the impedance of a cubic centimeter of tissue at the frequency $f$ in Hz.

$X_P$ in $\Omega/\text{cm}^3$ is the specific polarization reactance at the frequency $f$.

$j$ is an operator indicating that the current is shifted by a quarter period ahead of the voltage.

$R$ in $\Omega/\text{cm}^3$ is the specific resistance of the protoplasm and intercellular spaces in the tissue.

$r$ in $\Omega/\text{cm}^3$ is the ohmic resistance of the membranes in the tissue cube.

$C_1$ in F/cm$^3$ is the membrane capacitance in the tissue cube at the unity frequency.

$\beta$ in $\Omega/\text{cm}^3$ is the polarization reactance of the cell membranes in the tissue cube at the unity frequency. The last two parameters have an inverse relationship as in

$$\frac{1}{C_1} = \beta \ .$$

$\alpha$ is a numerical coefficient indicating the rate of decrease in the polarization reactance of the capacitor as a function of the frequency according to (7).

4) The parameter $R$ characterizes the physio-chemical state of protoplasm. The parameters $r$, $C_1$, $\alpha$, and $\beta$ characterize the physiological state of the membranes.

5) With changes in the physiological state of the tissue, these parameters can vary independently.